\documentclass[journal,twocolumn]{IEEEtran}
\usepackage[dvips]{graphicx}

\newcommand{\bstfile}{IEEE} 

\newcommand {\beq}{\begin{equation}}
\newcommand {\eeq}{\end{equation}}
\newcommand {\beqa}{\begin{eqnarray}}

\newcommand {\eeqa}{\end{eqnarray}}

\newcommand{\eqref}[1]{(\ref{#1})}

\begin{document}
\title{A general theory of inhomogeneous broadening for nonlinear susceptibilities: the polarizability and hyperpolarizability}
\author{Robert~J.~Kruhlak, and Mark~G.~Kuzyk
\thanks{Robert J. Kruhlak is at the Department of Physics, University of Auckland, Auckland, NZ E-mail: r.kruhlak@auckland.ac.nz}%
\thanks{Mark G. Kuzyk is at the Department of
Physics, Washington State University Pullman, Washington 99164-2814
E-mail: kuz@wsu.edu}}


\maketitle
\begin{abstract}
While nonlinear optical spectroscopy is becoming more commonly used
to study the excited states of nonlinear-optical systems, a general
theory of inhomogeneous broadening is rarely applied in lieu of
either a simple Lorentzian or Gaussian model.  In this work, we
generalize all the important linear and second-order nonlinear
susceptibility expressions obtained with sum-over state quantum
calculations to include Gaussian and stretched Gaussian
distributions of Lorentzians.  We show that using the correct model
to analyze experiments that probe a limited wavelength range can be
critical and that this theory is better able to fit the subtle
spectral features - such as the shoulder region of a resonance -
when both models produce qualitatively similar responses.
\end{abstract}



\section{Introduction}

\IEEEPARstart{W}{hen} calculating the nonlinear-optical
susceptibilities near resonance, it is important to account for
damping to get a reasonable approximation to the dispersion of a
real molecule. Clearly, neglect of damping is catastrophic since
sum-over-states (SOS) theories predict an infinite susceptibility on
resonance.  The SOS expression, as derived by Orr and
Ward,\cite{orr71.01} naturally admit damping corrections by making
the eigenenergies of the system complex. The imaginary part of the
energy is related to the resonance width and is called the natural
linewidth.  A stationary isolated molecule's line-shape is thus
described by a Lorentzian. When the width is related to the
intrinsic properties of a molecule, the width of a peak for a
collection of  noninteracting molecules is referred to as
homogeneous broadening, i.e. the observed spectrum is the same as
that of a single molecules times the number of molecules.

If the molecules interact or are placed in an environment with
random perturbations, the natural linewidth is smeared out, leading
to what is called inhomogeneous broadening (IB).  Stoneham described
in a detailed review article how the linear absorption spectrum in a
crystal could be accurately modeled by applying stochastic averaging
to a distribution of Lorentzians.\cite{stoneh69.01}  Since such
statistical processes often lead to gaussian peak shapes such as the
Boltzmann velocity distribution, systems with resonance features
whose widths are much larger than the natural line width are often
modeled by Gaussian functions.

Dirk et al\cite{dirk.90.04} and Berkovic et al\cite{berko00.01}
showed that it was important to take damping into account by using a
complex energy in the Orr and Ward SOS expression even when
considering off-resonance susceptibilities. The distribution of
molecules, when embedded in a polymer, are randomly perturbed by
virtue of the fact that each molecule experiences a different local
environment due to microscopic inhomogeneities.  The distribution of
sites around a molecule in a polymer can be
measured\cite{ghebr95.04,ghebr97.01} and quantified by a stretched
exponential distribution function.

Many excited states are often needed to correctly describe the
dispersion of the hyperpolarizability\cite{Dirk89.01,champ06.01} and
vibronic overtones may complicate the
analysis.\cite{wang00.01,wang01.01} Since it has been shown that
vibronic states are unimportant
off-resonance,\cite{Tripa04.01,tripa07.01} and on-resonance, we
assume that their affect is smeared out due to thermal broadening
(i.e. thermal fluctuations and the distribution of sites cause
variations that are larger than vibronic energies), we argue that
neglect of vibronic states and applying a stochastic averaging over
Lorentzian functions using a stretched exponential weighting
distribution function leads to a good model of the nonlinear
spectra.

In this work, we focus on using the SOS expression to calculate the
linear and second-order nonlinear-optical response under
inhomogeneous broadening using the above approximations.  (We will
apply the same approach to the dipole-free SOS
expression\cite{kuzyk05.02} in a future publication.)  Our
theoretical results are compared with experimental data in dye doped
polymers for validation.

\section{Theory}

Sum-over states quantum perturbation treatments of the $b^{th}$-order nonlinear susceptibility
tensor, $\xi_{ij...k}^{(b)}$, in the dipole approximation yields a sum of terms of the form:\cite{orr71.01}
 \beqa
 \lefteqn{\xi_{ij...k}^{(b)} \propto \sum_{n}^{\infty} \sum_{m}^{\infty}
  ...\sum_{l}^{\infty}} \nonumber \\
  &&\hspace{-0.3in} \frac {\left(\mu_{i}\right)_{gn} \left(\mu_{j}\right)_{nm}
  ... \left(\mu_{k}\right)_{lg}} {(\omega_{ng} - i \Gamma_{ng} -  \omega_1)
   (\omega_{mg}-i \Gamma_{mg} -  \omega_1 -  \omega_2) ...} ,
 \label{general} 
 \eeqa
where $\left( \mu_{i} \right)_{nm}$ is the $nm$-matrix element of
the $i^{th}$ component of the electric dipole operator,
$\omega_{nm}$ the transition frequency (energy) between states $n$
and $m$, $\omega_i$ the frequency of the $i^{th}$ optical field, and
$\Gamma_{ng}$ the phenomenological damping factor.  The numerator is
a product of $b+1$ transition moments and the denominator a product
of $b$ energy terms.  For an isolated molecule, the damping factor
$\Gamma_{ng}$ is inversely proportional to the lifetime of state $n$
and is a measure of the width of the peak in the spectrum of
$\xi_{ij...k}^{(b)}$ associated with a transition between state $n$
and the ground state $g$.

In real systems, molecules interact with each other yielding a broadening of the peaks in a
spectrum.  One common method to treat this case is to adjust the parameters $\Gamma_{ng}$.  If the
statistics of the molecules are Gaussian (as one finds in Doppler broadening), the Lorentzian
function resulting from adjusting $\Gamma_{ng}$ has the correct width but does not have a Gaussian
shape.  Another common method for modeling a nonlinear spectrum is to assume the peaks are
Gaussian in shape.  Depending on the region of interest (i.e near the peak or in the tail of the
spectrum) one model may be more applicable than the other.

An exact method of treating inhomogeneous broadening is to apply the statistics of the molecular
interactions to the susceptibility.  Each molecule in an ensemble is then viewed as having a
different transition frequency (energy), $\omega_{ng}$.  For example, if the statistics are
Gaussian with a probability distribution of the form
 \begin{equation}
 f_{ng}(\delta\omega_{ng})=\frac {1} {N (\gamma_{ng})}
  \exp \left[-  \left( \frac {\delta \omega_{ng}} { \gamma_{ng}} \right)^2 \right],
   \label{distribution}
 \end{equation}
where $\delta \omega_{ng}= \omega_{ng} - \bar{\omega}_{ng}$, $\bar{\omega}_{ng}$ is the mean value
of the transition frequency, $N(\gamma_{ng})$ the normalization factor, and $\gamma_{ng}$ the
linewidth of the distribution - the susceptibility will be of the form,
  \beqa
 \lefteqn{\left( \int_{- \bar{\omega}_{ng}}^\infty d(\delta \omega_{ng}) \int_{- \bar{\omega}_{mg}}^\infty
  d(\delta \omega_{mg}) ... \right)} \nonumber \\
 && \xi_{ij...k}^{(b)} (\omega_{ng}, \omega_{mg},...) f_{ng}(\delta\omega_{ng}) f_{mg}(\delta\omega_{mg}) ...
  \label{convolution} 
 \eeqa
Toussaere developed such a theory for second harmonic generation
(SHG) and the Pockels effect using Gaussian
statistics.\cite{touss93.01}

In materials such as dye-doped polymers, the statistics that best model the effect of the distribution of sites on processes such as relaxation of molecular orientation order (which is normally an exponential process as described by Debye) is a stretched exponential.  Using this analogy, we propose that variations in the local electric fields in a polymer yields the same statistics for modeling the susceptibility, so the distribution is also assumed to be a stretched Gaussian of the form,
 \begin{equation}
 f_{ng}(\delta\omega_{ng})=\frac {1} {N(\gamma_{ng},\beta)}
   \exp \left[ - \left( \frac {\delta\omega_{ng} }
    {\gamma_{ng}} \right)^{2\beta} \right] ,
  \label{stretched}
 \end{equation}
where $\beta$ is the distribution of sites parameter, and for many systems - such as in dye-doped polymers - varies from $\beta=0$ for an infinitely
broad distribution to $\beta =1$ for a single characteristic width.  Again, $N(\gamma,\beta)$ is a
normalization factor, which depends also on $\beta$, and will be written as
 \beq
 N(\gamma_{ng},\beta) = \gamma_{ng} \sqrt{\pi} B(\beta),
 \label{Ngambet}
 \eeq
 where,

 \beq
 B(\beta) =  \left[ \frac{1}{\gamma_{ng} \sqrt{\pi}} \int^{\infty}_{-\infty}
               \exp \left[ - \left( \frac {\delta\omega_{ng} }
    {\gamma_{ng}} \right)^{2\beta} \right] d(\delta\omega_{ng})\right]
 \label{Bbeta}
 \eeq
to remain compatible with previous  inhomogeneous broadening representations which use Gaussian
statistics \cite{touss93.01,otomo95.01,kruhl99.01,kruhl99.02}. Indeed, such statistics other than
pure Gaussian were observed in light scattering experiments \cite{kruhl97.01,kruhl99.01,kruhl99.02}.

\subsection{Normalization and Approximations}

\begin{figure}[!h]
 \begin{center}
 \scalebox{.4}{\includegraphics{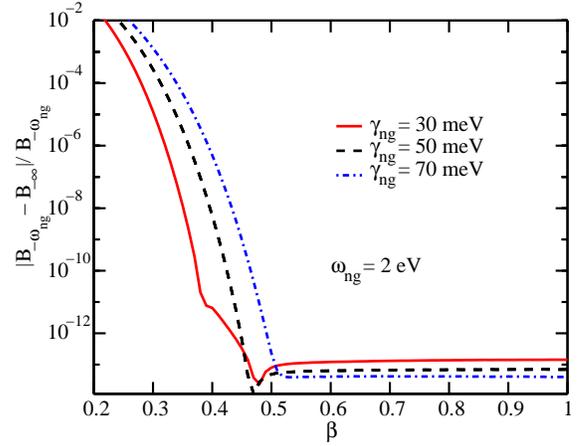}}
 \end{center}
 \caption{{Relative difference between $B(\beta$) when the integral is taken over
  all frequency space, and when the lower limit for $B(\beta$) is given by $\omega_{ng} = 2$ eV}}
  \normalsize
  \label{fig:Blowlimitvsbeta}
  \end{figure}
Because $B(\beta)$ is difficult to calculate, we need to make
simplifying assumptions.  Typical inhomogeneous linewidths are
narrow compared with the transition frequency \cite{kruhl00.01,
kruhl99.02}, so we can integrate over all space even though there is
a finite lower limit to the integral.  Furthermore, we assume that
$B(\beta$) is independent of $\gamma_{ng}$. Figure
\ref{fig:Blowlimitvsbeta} shows the relative difference between
$B(\beta$) determined by integrating of over all frequency space and
integration to a lower limit of $\omega_{ng}= 2$ eV. Clearly, if the
inhomogeneous linewidth is smaller than 70 meV -- which is the case
for all our experimental data -- this assumption leads to a
negligible relative difference ($< 1 \%$) for $\beta > 0.3$.
\begin{figure}[!h]
 \begin{center}
 \scalebox{.4}{\includegraphics{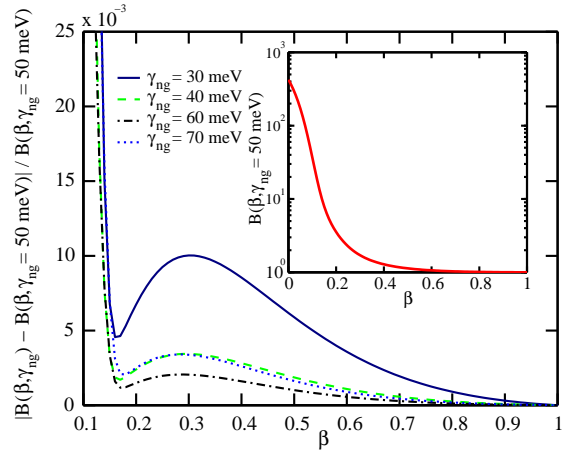}}
 \end{center}
 \caption{{Relative difference between B($\beta, \gamma_{ng}$) and
  B($\beta, \gamma_{ng}$ = 50 meV) }}
  \normalsize
  \label{fig:Brelerrorvsgamma}
  \end{figure}

The second assumption is that the normalization function $B(\beta$)
is independent of the inhomogeneous linewidth, $\gamma_{ng}$. Figure
\ref{fig:Brelerrorvsgamma} shows that this assumption is valid to
good approximation (within 1\%) for $\beta > 0.15$ (see inset in
Figure \ref{fig:Brelerrorvsgamma}).  The reference value of
$B(\beta, \gamma_{ng}= 50$ meV) was chosen because it is a typical
linewidth for the chromophores we consider. These assumptions allow
us to calculate one normalization function independent of
$\omega_{ng}$ and $\gamma_{ng}$ when we consider $\beta > 0.3$,
which is critical for decreasing the computation time of the theory
when applied to a least-squares fit to the data.

In this paper, we derive the expressions for the most important
linear and second-order susceptibilities for any general statistics.
Specific results are presented for Gaussian and stretched
exponential suceptibilities that model the SHG response
(hyperpolarizability) for a two-level system and  data from linear
absorption experiments (polarizability) on silicon
phthalocyanine-(mono)methylmethacrylate(SiPc) doped
(poly)methylmethacyrlate (PMMA) thin films. All the expressions for
the second-order processes are displayed in detail in Appendix B.

\subsection{First-Order Susceptibility}

In this section, we outline the approach of the calculation for a general distribution function
for a linear susceptibility (which has one energy denominator) and in the next section apply the
theory to a second-order susceptibility (which has two energy denominators).  These two cases can
be easily generalized to any-order susceptibility and any statistics.

From time dependent perturbation theory, the 1-D first-order
molecular susceptibility is \cite{butch90.01,kruhl00.01}: \beq
 \xi^{(1)}(-\omega; \omega) 
  =\frac{1}{ \epsilon_0 1!}\frac{1}{ \hbar}
 \sum_n
 \left \{ \mu_{gn}\mu_{ng}D^L_n(-\omega; \omega)\right\},
 \label{xi1_lorentz}
 \eeq
where
 \beq
 D^L_n(-\omega; \omega)=\left \{ \frac{1}{\Omega_{ng} -\omega} +
           \frac{1}{\Omega_{ng}^{*} +\omega} \right\}.
 \label{energydenomL1}
 \eeq
The inhomogeneous broadened Lorentzian function is then
  \beq
 D_n^{IB}(-\omega; \omega) = \int^{\infty}_{-\omega_{ng}} D^L_n(-\omega; \omega)
 f_{ng}(\delta\omega_{ng})d(\delta\omega_{ng}).
 \label{Dn_ib_integral}
 \eeq
Once $D_n^{IB}(-\omega; \omega)$ is known it can be substituted into
Eq. (\ref{xi1_lorentz}) for $D_n^L(-\omega;\omega)$ to create an
inhomogeneously broadened microscopic susceptibility. To this end,
we proceed to determine the first term in Equation
(\ref{Dn_ib_integral}).  Substituting Equation (\ref{stretched}) for
the Gaussian function, changing the integration variable to
$t=(\omega^{\prime}_{ng} - \omega_{ng})/\gamma_{ng}$, and
rearranging the denominator so that $z = (-\omega_{ng}+ i\Gamma_{ng}
+\omega)/{\gamma_{ng}}$ gives the following
 \beqa
 \lefteqn{\int^{\infty}_{-\omega_{ng}} \frac{1}{\omega'_{ng} -i\Gamma_{ng} -\omega}
f_{ng}(\omega'_{ng} -\omega_{ng}) d(\omega'_{ng} -\omega_{ng})} \nonumber\\
 &= & \frac{1}{\gamma_{ng} \sqrt{\pi} B(\beta)}
 \int^{\infty}_{-\frac{\omega_{ng}}{\gamma_{ng}}} \frac{\exp({-t^{2\beta}})}
 {t + (\frac{\omega_{ng} -i\Gamma_{ng}-\omega}{\gamma_{ng}})} dt, \nonumber \\
&\simeq&\hspace{-0.09 in} -\frac{1}{\gamma_{ng} \sqrt{\pi} B(\beta)}
\int^{\infty}_{-\infty} \frac{\exp({-t^{2\beta}})}{z -t} dt.
  \label{edfotrans}
 \eeqa
To simplify the expression in Equation \eqref{edfotrans}, we define $W^{(k)}_{\beta}(z)$ as the
following:
 \beq
 W^{(k)}_{\beta}(z)= \frac{i}{\pi B(\beta)}\int^{\infty}_{-\frac{\omega_{ng}}{\gamma_{ng}}}
  \frac{\exp({-t^{2\beta}})}{(z-t)^{k}} dt,
 \label{WWb}
 \eeq
 and the complex error function as \cite{abram72.01},
 \beq
 W(z)= \frac{i}{\pi}\int^{\infty}_{-\infty} \frac{\exp({-t^2})}{z-t} dt
 \label{WW}
 \eeq

 The second term in Eq. (\ref{Dn_ib_integral}) is derived analogously so we write the first-order energy denominator that accounts for inhomogeneous
 broadening as:
  \beqa
 \lefteqn{D_n^{IB}(-\omega; \omega)=\frac{i\sqrt{\pi}}{\gamma_{ng}} \times}
  \\
 &&\left[ W^{(1)}_{\beta}\left(\frac{ -(\Omega_{ng} - \omega)}{\gamma_{ng}} \right) +
  W^{(1)}_{\beta}\left(\frac{ -(\Omega^{*}_{ng} + \omega) }{\gamma_{ng}} \right)
  \right].\nonumber
 \label{Dnibgen}
 \eeqa
 In the limit that $\beta = 1$, the first-order energy denominator can be written in terms of the complex error function,
 $W(z)$ \cite{touss93.01,otomo95.01,kruhl99.01}:
 \beqa
 \lefteqn{ D_n^{IB}(-\omega; \omega)=\frac{i\sqrt{\pi}}{\gamma_{ng}}
 \times } \\
 &&\left[ W\left(\frac{ -(\Omega_{ng} - \omega)}{\gamma_{ng}} \right) +
  W\left(\frac{ -(\Omega^{*}_{ng} + \omega)}{\gamma_{ng}} \right) \right], \nonumber
 \label{Dnib}
 \eeqa
which can be evaluated for all values of $z$ using the results in
Abramowitz and Stegun \cite{abram72.01}. The general transformation
from the Lorentzian  model to the inhomogenous broadening model is
summarized in Table \ref{tab:transforms}.

\subsection{Second-Order}

Using time dependent perturbation theory, we define a second-order
Lorentzian energy denominator as:
\beqa
 \lefteqn{D^L_{lm}(-\omega_{\sigma}; \omega_1, \omega_2)  = {\mathbf{S}}_{1,2}
 \left\{ \left[ (\Omega_{lg} - \omega_{\sigma}) (\Omega_{mg} - \omega_1) \right]^{-1}
 \right.  } \nonumber \\
 && \hspace{0.7in}+ \left[ (\Omega_{lg}^{*} + \omega_2)(\Omega_{mg}^{*} + \omega_{\sigma}) \right]^{-1}
  \nonumber \\
 &&  \hspace{0.7in} \left.
   + \left[ (\Omega_{lg}^{*}+\omega_2)(\Omega_{mg} - \omega_1)\right]^{-1}
  \right\}, \label{Dlm}
 \eeqa
where ${\mathbf{S}}_{1,2}$ is the permutation operator, which
averages over all distinct permutations of $\omega_1$, and
$\omega_2$.

It is significantly more difficult to transform nonlinear Lorentzian
energy denominators to nonlinear IB energy denominators because of
the product of Lorentzian terms in the denominator.   In order to
transform $D^L_{lm}$ for a specific experiment (i.e. for specific
input and output frequencies), the number of excited states must be
known prior to performing a partial fraction expansion on each term
in the energy denominators. For example if there is only one
distinct excited state ($l$), it maybe necessary to perform the
following partial fraction expansion,
 \beq
  \frac{1}{(\Omega_{lg}-\omega)\Omega_{lg}}=
  \frac{1}{\omega}\left[\frac{1}{(\Omega_{lg}-\omega)} -
                        \frac{1}{\Omega_{lg}}\right],
 \eeq
to eliminate the product of the two $\Omega_{lg}$ terms. These type
of expansions allow us to write the nonlinear energy denominators in
terms of $W^{(1)}_{\beta}(z)$ or complex error functions.

However, a perfect square in the denominator requires a different
approach.  The approach for evaluating the fundamental
transformations from the homogeneous formulation to the
inhomogeneous formulation for the quadratic dependencies (cubic
dependencies are studied in Ref. \cite{kruhl08.02}) on the
transition frequency, $\omega_{lg}$ is shown below.

Beginning with a quadratically dependent term like the following,
 \beq
 \frac{C_2}{(\omega'_{ng} -i\Gamma_{ng} -\omega)^2}\label{edsoexamp}
 \eeq
 we  integrate its product with the distribution function (Equation \eqref{stretched}),
 \beq
 \int^{\infty}_{-\omega_{ng}} \frac{C_2}{(\omega'_{ng} -i\Gamma_{ng} -\omega)^2}
 f_{ng}(\omega'_{ng} -\omega_{ng})\,d(\omega'_{ng} -\omega_{ng}), \label{spso}
 \eeq
as the initial step in the transform. Substituting Equation (\ref{stretched}) for
$f_{ng}(\omega'_{ng} -\omega_{ng})$, and changing the integration variable to
$t=(\omega^{\prime}_{ng} - \omega_{ng})/\gamma_{ng}$, results in the following:
 \beqa
 \lefteqn{ \int^{\infty}_{-\omega_{ng}} \frac{C_2}{(\omega'_{ng}
-i\Gamma_{ng} -\omega)^2} f_{ng}(\omega'_{ng} -\omega_{ng}) d(\omega'_{ng} -\omega_{ng})}
\nonumber
\\
 &=&\hspace{-0.09 in} \frac{C_2}{\gamma_{ng} \sqrt{\pi} B(\beta)} \int^{\infty}_{-\omega_{ng}}
  \frac{\exp({-(\frac{\omega'_{ng}-\omega_{ng}}{\gamma_{ng}})^{2\beta}})}
  {(\omega'_{ng} -i\Gamma_{ng} -\omega)^2} d(\omega'_{ng} -\omega_{ng}), \nonumber\\
 &= &\hspace{-0.09 in} \frac{C_2}{\gamma_{ng} \sqrt{\pi} B(\beta)}
 \int^{\infty}_{-\frac{\omega_{ng}}{\gamma_{ng}}}
 \frac{\gamma_{ng}\exp({-t^{2\beta}})}{(\omega'_{ng} -i\Gamma_{ng}-\omega)^2} dt,\nonumber\\
 &= & \hspace{-0.09 in}
 \frac{C_2}{\gamma_{ng} \sqrt{\pi} B(\beta)} \int^{\infty}_{-\frac{\omega_{ng}}{\gamma_{ng}}}
 \frac{\gamma_{ng}\exp({-t^{2\beta}})}{(\omega_{ng} + \gamma_{ng}t -i\Gamma_{ng}-\omega)^2} dt,
 \nonumber
 \\ &= &\hspace{-0.09 in}
 \frac{C_2}{\gamma_{ng} \sqrt{\pi} B(\beta)} \int^{\infty}_{-\frac{\omega_{ng}}{\gamma_{ng}}}
 \frac{\gamma_{ng}\exp({-t^{2\beta}})}{\gamma_{ng}^2 ( t +\frac{
 \omega_{ng}-i\Gamma_{ng}-\omega}{\gamma_{ng}})^2} dt,\nonumber \\
 &= &\hspace{-0.09 in} \frac{C_2}{\gamma_{ng}^2 \sqrt{\pi} B(\beta)}
  \int^{\infty}_{-\frac{\omega_{ng}}{\gamma_{ng}}} \frac{\exp({-t^{2\beta}})}{ (z - t)^2} dt,
  \nonumber \\
 &= &\hspace{-0.09 in} \frac{C_2 i \sqrt{\pi}}{\gamma_{ng} }
 \left[\frac{-1}{\gamma_{ng}}W^{(2)}_{\beta}(z)\right]
 \label{edsochvar}
 \eeqa
where $z=(- \omega_{ng}+ i\Gamma_{ng}+\omega)/{\gamma_{ng}}$. In the limit that $\beta=1$,
Equation (\ref{edsochvar}) looks very similar to Equation (\ref{edfotrans}) just before
substituting for $W(z)$ except that the denominator in the integral is second-order in $(z-t)$. To
reduce the denominator to first-order in $(z-t)$ so that the integral can be replaced with $W(z)$,
we perform integration by parts twice:
 \beqa
 \int^{\infty}_{-\frac{\omega_{ng}}{\gamma_{ng}}} \frac{T}{ (z - t )^2} dt
 &=&  \left. \frac {T} {z-t} \right|_{t = - \frac {\omega_{ng}} {\gamma_{ng}}}
  \hspace{-.75in}
  {\raisebox{-2ex}{\mbox{
   {\includegraphics{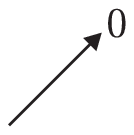}}
  }}}
  \hspace{+.25in}
 + \int^{\infty}_{-\frac{\omega_{ng}}{\gamma_{ng}}} \frac{2tT}{ (z - t )} dt
 , \nonumber \\
 &=& 2z \int^{\infty}_{-\frac{\omega_{ng}}{\gamma_{ng}}} \frac{T}{ (z - t )} dt
 -2\int^{\infty}_{-\frac{\omega_{ng}}{\gamma_{ng}}}T dt
 , \nonumber \\
 &\simeq& -2 i \pi z W(z) -2 \sqrt{\pi}
 ,\label{quadi}
 \eeqa
where we have used $T=\exp({-t^2})$ to simplify the presentation.
The first term on the right on the top line of Equation
(\ref{quadi}) vanishes because the argument of the exponential is
$\approx - 10^3$ at the lower limit ( usually $\omega \gg \gamma$ in
the visible).  Note that Equation \eqref{quadi} is kept general (the
lower limit of integration is not approximated as $-\infty$). We
only use this approximation as needed for special cases.  To get the
second line of Equation (\ref{quadi}), we use $t/(z-t) = z/(z-t) -
1$. Therefore we can write Equation (\ref{spso}), when $\beta=1$, as
the following,
 \beqa
 \lefteqn{\int^{\infty}_{-\omega_{ng}} \frac{C_2}{(\omega'_{ng} -i\Gamma_{ng} -\omega)^2}
  f_{ng}(\omega'_{ng} -\omega_{ng}) d(\omega'_{ng} -\omega_{ng}) } \nonumber \\
& =&
  \frac{C_2}{\gamma_{ng}^2 \sqrt{\pi}}\left\{ -2 i \pi z W(z) -2 \sqrt{\pi}
  \right\} \hspace{1.3in} \nonumber\\
 & =&
 \frac{C_2 i  \sqrt{\pi}}{\gamma_{ng}}\left \{ \frac{ -2 z}{\gamma_{ng}} W(z)
  + \frac{2 i}{ \sqrt{\pi} \gamma_{ng}}
  \right \}. 
   \label{spsof}
 \eeqa

Thus we have derived the convolution of a second-order denominator
term with a Gaussian, which can be generalized to any complex term
that has a second-order dependence on $(z-t)$.

Table \ref{tab:transforms} in Appendix A summarizes the fundamental
energy denominators up to second-order for the Lorentzian and IB
theories with $\beta \leq 1$, and $\beta =1$, respectively. These
two transformations can be used to construct any IB energy
denominator for any first-, and/or second-order response.

\begin{table}[!h]
\begin{center}
 \caption{{\label{tab:compact} Compact form of $W^{(x)}_{\beta}(z)$ and $W(z)$.} }
\begin{tabular}[!h]{ccc}
$\beta$ & Term & Compact Form \\\hline
 $\leq 1$& $
  W^{(x)}_{\beta}\left(\frac{-(\Omega_{ng} \mp
\omega_i)}{\gamma_{ng}}\right) $ & $W^{(x)}_{\beta_n}(\mp\omega_i) $\\
& $W^{(x)}_{\beta}\left(\frac{-(\Omega^*_{ng} \pm
\omega_i)}{\gamma_{ng}}\right)$ &
$W^{(x)^*}_{\beta_n}(\pm\omega_i)$\\
  &&\\
 $1$& $W\left(\frac{-(\Omega_{lg} \mp \omega_i)}{\gamma_{lg}}\right)$ & $W_l(\mp\omega_i)$ \\
&$W\left(\frac{-(\Omega^*_{lg} \pm \omega_i)}{\gamma_{lg}}\right)$ & $W^*_l(\pm\omega_i)$\\
  \hline \\
\end{tabular}
\end{center}
 \end{table}

The energy denominators $D_{lm...}$ for the second-(and
higher)-order susceptibilities are complex combinations of
$W^{(x)}_{\beta}(z)$ or $W(z)$. To make them more readable we have
developed a compact notation. For $\beta \leq 1$, we have added a subscript to
$\beta$ to indicate the excited state involved in the process and a
$^*$ on the power of $W$ to indicate a complex conjugate of the
complex argument $\Omega_{ng} = \omega_{ng} - i\Gamma_{ng}$.
Similarly for $\beta=1$, the subscript on $W$ indicates the excited
state and the superscript $^*$ on $W$ indicates the complex
conjugate of $\Omega$. This allows us to use a simple frequency
argument of $\pm \omega$ which significantly improves the
readability of the equations in Appendix B. An example of the
transformation to the compact notation is given below: \beq
W^{(1)}_{\beta}\left(\frac{-(\Omega^*_{2g} +
\omega_3)}{\gamma_{2g}}\right) \rightarrow
W^{(1)^*}_{\beta_2}(\omega_3) \eeq This compact form is sufficient
to describe  all of the inhomogeneous broadening contributions to
the energy denominators in this paper because the arguments are all
of the  form $\frac{-(\Omega^*_{ng} \pm \omega_i)}{\gamma_{ng}}$ or
$\frac{-(\Omega_{ng} \mp \omega_i)}{\gamma_{ng}}$ and all of the
excited state transitions are from/to the ground state ($g$). Table
\ref{tab:compact} summarizes the transformations to the compact
notation.
\begin{figure}[!ht]
\begin{center}
\scalebox{1}{\includegraphics{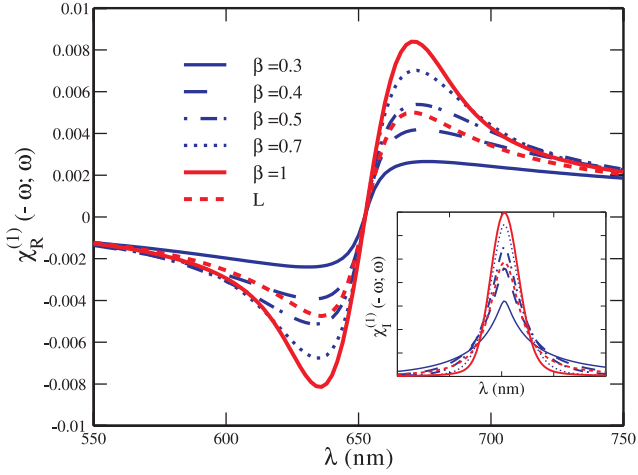}}
\end{center}
\caption{{Real and imaginary [inset] parts of $\chi^{(1)}(-\omega;
\omega)$ for a single excited state centered at 653 nm for various
values of $\beta$. For all values of $\beta$, $\Gamma = 10$ meV and
$\gamma = 50$ meV. The homogeneous-broadening (Lorentzian) theory of
the electronic transition is denoted by L and $\Gamma = 50$ meV.}}
\normalsize \label{fig:chi1vsbeta}
 \end{figure}

\subsection{Linear Susceptibility}

\begin{figure}[!htb]
\begin{center}
\scalebox{0.45}{\includegraphics{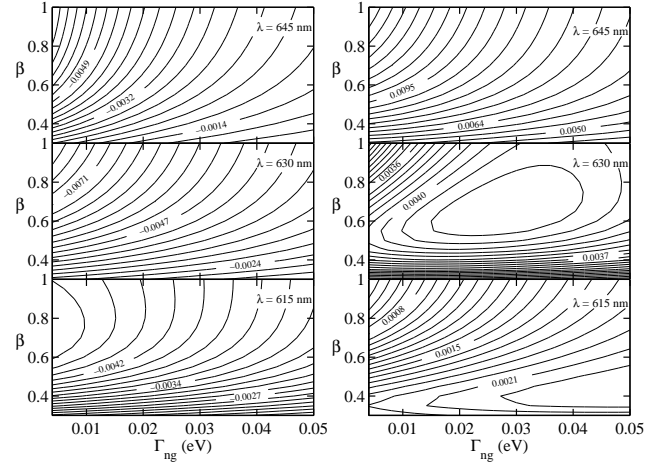}}
\end{center}
\caption{{Contours of the real and imaginary parts of $\chi^{(1)}(-\omega; \omega)$ generated
using the generalized IB theory for a single excited state centered at 653 nm ($\mu_{ng} = 11.5$
D, $\gamma_{ng}= 50 $ meV) for three probe wavelengths (615 nm, 630 nm, 645 nm).}}
 \normalsize \label{fig:chi1betaLorentz}
 \end{figure}

In this section we discuss the effects of $\beta$ on the linear response of a
dye-doped polymer system with one excited state. Figure
\ref{fig:chi1vsbeta} shows the real and imaginary parts of the
linear susceptibility, which is proportional to Eq.
(\ref{xi1_lorentz}) with $n=1$ and the appropriate energy
denominators, for various values of $\beta$ in the IB model. For
comparison the homogeneous (Lorentzian) theory is denoted by L. We
use typical values for the transition moment and homogeneous and
inhomogeneous linewidths. In this example the excited state is
centered at about 653 nm with a transition moment of 11.5 D. For all
the IB curves the Lorentzian linewidth is 10 meV, and the IB
linewidth is 50 meV. The Lorentzian curve is generated using a 50
meV linewidth and the same center frequency and transition moment as
the IB curves. As we expect the magnitude of the response for a
fixed inhomogeneous linewidth decreases and becomes broader as the
distribution of sites within the polymer matrix increases
(decreasing $\beta$).
\begin{figure}[!htb]
\begin{center}
\scalebox{0.45}{\includegraphics{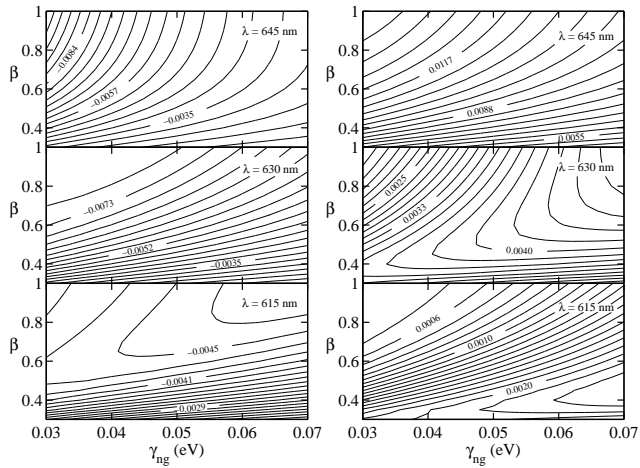}}
\end{center}
\caption{{Contours of the real (left) and imaginary (right) parts of
$\chi^{(1)}(-\omega; \omega)$ generated using the generalized IB
theory for a single excited state centered at 653 nm ($\mu_{ng} =
11.5$ D, $\Gamma_{ng}= 10 $ meV) for three probe wavelengths (615
nm, 630 nm, 645 nm). }}
 \normalsize \label{fig:chi1betagamma}
 \end{figure}
To better describe the effects of the distribution of sites, IB,
and homogenous broadening in the linear response we plot the real
and imaginary parts of $\chi^{(1)}(-\omega; \omega)$ as a function
of $\beta$ (distribution of sites) and $\gamma_{ng}$ (IB) for fixed
$\Gamma_{ng}$(L) and $\chi^{(1)}(-\omega; \omega)$ as a function of
$\beta$  and $\Gamma_{ng}$  for fixed $\gamma_{ng}$. In Figure
\ref{fig:chi1betagamma}, the single excited state transition is
centered at 653 nm, with a transition dipole moment of 11.5 D, and a
fixed Lorentzian linewidth of 10 meV, while Figure
\ref{fig:chi1betaLorentz} uses a fixed IB linewidth, $\gamma_{ng}$
of 50 meV.

Figures \ref{fig:chi1betagamma} and \ref{fig:chi1betaLorentz} each
show a distinct character of $\chi^{(1)}(-\omega; \omega)$ for
wavelengths between 615 nm and 645 nm.  The $\beta$ parameter
therefore gives an additional degree of freedom for modeling
guest/host materials where a distribution of sites requires using
non-Gaussian statistics for both the index of refraction and the
absorption coefficient.

\section{Second-order Susceptibility}
\begin{figure}[!ht]
\begin{center}
\scalebox{0.85}{\includegraphics{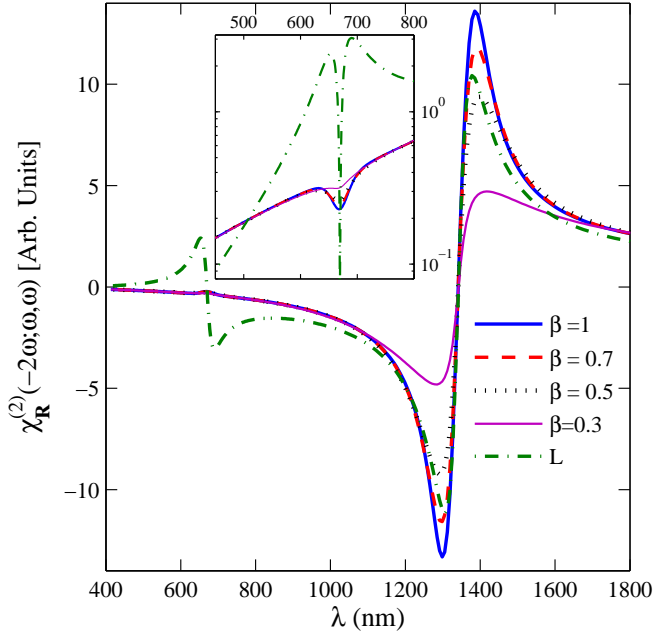}}
\end{center}
\caption{{Second harmonic generation response for several
inhomogeneously broadened two level systems. The real part of
$\chi^{(2)}(-2\omega;\omega,\omega)$ is shown as a function of the
incident wavelength. A homogenously broadened (L) response is
included for comparison. The inset shows the magnitude on a log
scale in the visible region. Parameters for the curves can be found
in Table \ref{tab:chi2param}.}}
 \normalsize \label{fig:shgtwolevel}
 \end{figure}
In this section, we demonstrate that there are significant
differences in the response from  two-level systems depending on the
broadening model. We concentrate on second
harmonic generation (SHG) as an example because of its
near-universal use in characterizing molecules.
For a two-level system, the SHG hyperpolarizability
$\chi^{(2)}(-2\omega; \omega,\omega) \propto D_{ll}(-2\omega;
\omega,\omega)$ [see Eq (\ref{Dllshgibbetashort}) and Eq.
(\ref{Dllshgibshort}) for the IB models, and Eq. (\ref{Dlm}) for the
L model] where $l$ in this case is 1.

\begin{table}[!h]
\begin{center}
 \caption{{\label{tab:chi2param} Inhomogeneous and Lorentzian broadening parameters for a two-level system.} }
\begin{tabular}[!h]{cccc}
\hline \\
 Model&$\omega_{1g}$ &$\Gamma_{1g}$& $\gamma_{lg}$ \\\hline
 L& 1.85 eV& 0.05 eV & -- \\
 IB& 1.85 eV& 0.025 eV & 0.05 eV \\
  \hline \\
\end{tabular}
\end{center}
 \end{table}

Figure \ref{fig:shgtwolevel} shows the response of a
two-level system using the parameters in Table
\ref{tab:chi2param}. For the IB model four values of $\beta$ have
been studied as shown in the legend. It is clear that the
response in the IR for all models are qualitatively similar but
quantitatively different. Thus a fit to an SHG experiment that
probes this system in the IR region would come to a similar
conclusion for the parameters describing which excited states are involved
in the process. The difference between the two models could be used
to estimate the error.

In the visible part of the spectrum (see inset), the different models
do not agree with each other, where the L model is dramatically different both qualitatively and
quantitatively from the IB models. This becomes
important for those experiments that probe the system
over a limited wavelength range (or more egregiously, at a single wavelength) and
attempt to determine an off-resonance value for the
hyperpolarizability.  In this example, an SHG experiment that probed
this one-level system only in the visible would determine a
completely different set of parameters depending on the model used
to interpret the results.

\section{Comparison of Theory to Experimental Results}
\begin{figure}[!ht]
 \begin{center}
 \scalebox{.4500}{\includegraphics{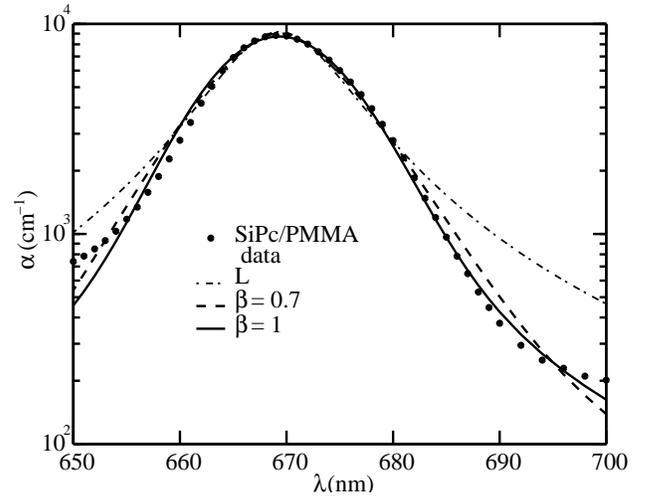}}
 \end{center}
 \caption{{Linear absorption of SiPc/PMMA in comparison to least-squares fits using Lorentzian
  and IB theories ( L=$\{\mu_{1g} = $ 8 D, and $\Gamma_{1g} =$ 18 meV  $\}$,
      IB($\beta = 1$) = $\{\mu_{1g} = $ 7.4 D, $\Gamma_{1g} =$ 9 meV, and $\gamma_{1g} =$ 19 meV $\}$,
      IB($\beta = 0.7$) = $\{\mu_{1g} =$ 7.5 D, $\Gamma_{1g} = $ 7 meV, and $\gamma_{1g} =$ 18 meV  $\}$). }}
  \normalsize
  \label{fig:alphavslamsipc}
  \end{figure}

Next, we compare IB Lorentzian models
with linear absorption data from (SiPc/PMMA) thin
films\cite{kruhl05.01} to show that the wing region of the linear
absorption resonance is often better described using an
inhomogeneous broadening model. A similar comparison for the
quadratic electroabsorption spectrum of this molecule can be found
in the literature.\cite{kruhl05.01,kruhl08.02}

Figure \ref{fig:alphavslamsipc} compares SiPc experimental linear
absorption data with the Lorentzian (L) theory, and the IB theory
for two values of $\beta$. A least squares fit gives $\beta=0.9$. It
is possible that the high-energy side of the peak is affected by the
next excited state, so a least squares fit was performed from the
peak to the low-energy side. This gives a least squares fit of
$\beta = 1$.  In either case, it appears that the distribution of
sites is narrow in the SiPc/PMMA system.  The result is reasonable
given that the SiPc molecule is non-dipolar so
there are no dipole interactions with the surroundings.  All
broadening therefore results from higher-order multipoles, which
contribute more weakly to the width.

Given that the calculated homogeneous broadened linewidths for each
model compare well with the literature
on the temperature dependence of these linewidths for doped PMMA
systems,\cite{garci97.01} it is clear that the IB model describes the response better
than the Lorentzian model especially in the wings of the resonance.
It should therefore give a better estimate of zero-frequency
susceptibilities for comparison with the Thomas--Kuhn sum-rule
quantum limit.\cite{kuzyk00.01,kuzyk03.01,kuzyk00.02,kuzyk03.02} Indeed,
it would be interesting to apply inhomogeneous broadening models to
the fundamental limits of the dispersion of the hyperpolarizability.\cite{kuzyk06.03}
It may very well be the case that the gap between the best nonlinear-optical
molecules and the fundamental limit\cite{Tripa04.01,tripa07.01,kuzyk03.02}
may be partially explained by inaccuracies in the dispersion model.

\section{Conclusion}

In conclusion, we calculate the inhomogeneously broadened linear and second-order nonlinear susceptibilities
for a Gaussian and stretched Gaussian distribution of Lorentzians.  A Lorentzian model is found to be inaccurate
in predicting the shape of the linear absorption spectrum of SiPc/PMMA.  However, the
IB-broadened theory fits the data over a broad range of wavelengths and shows that the
distribution of sites is nearly Gaussian, implying that interactions between the polymer and
dopant are small, as we would expect of a non-dipolar molecule.

Since the nature of the broadening mechanisms, especially for nonlinear spectroscopy, is shown to affect the dispersion
dramatically, the determination of excited state properties of molecules from spectroscopy
requires that such a theory be used.  Indeed, the inconsistency of transition moments as
determined by different processes (i.e. linear versus nonlinear spectroscopy) may be due in part
to the use of inappropriate dispersion models.  Furthermore, the practice of extrapolating single-wavelength
measurements to get the off-resonance hyperpolarizability, $\beta_0$,\cite{kuzyk98.01} may lead
to large inaccuracies.  As such, IB theory may be an important tool for interpreting any
nonlinear-optical experiment.

 \section*{Acknowledgments} We thank the National Science
Foundation (ECS-0354736) and Wright Patterson Air Force Base for generously supporting this work.
\bibliographystyle{\bstfile}

\appendices

 \onecolumn
\section{Energy denominator transformations}
\begin{table}[!hbt]
\begin{center}
 \caption{{\label{tab:transforms} Fundamental
energy (frequency) difference contributions to homogeneously
broadened (L) and inhomogeneously broadened (IB) electronic
transitions up to second-order.} } \footnotesize
\begin{tabular}[t]{ccc}   \hline
Lorentzian (L)&\multicolumn{2}{c}{Inhomogeneous Broadening (IB)} \\
&($\beta \leq 1$) &($\beta = 1$) \\
 \hline &&\\
 $\frac{C_1}{\omega_{ng} \mp i\Gamma_{ng}
\mp \omega}$ & $
\frac{i\sqrt{\pi}C_1}{\gamma_{ng}}W^{(1)}_{\beta}\left(\frac{-\omega_{ng}
\pm i\Gamma_{ng} \pm \omega}{\gamma_{ng}}\right)$& $
\frac{i\sqrt{\pi}C_1}{\gamma_{ng}}W\left(\frac{-\omega_{ng} \pm
i\Gamma_{ng} \pm
\omega}{\gamma_{ng}}\right)$\\ &\\
 $\frac{C_2}{(\omega_{ng} \mp i\Gamma_{ng} \mp \omega)^2}$ &
 $\frac{i\sqrt{\pi} C_2}{\gamma_{ng}} \left[ \frac{-1}{\gamma_{ng}}
 W^{(2)}_{\beta}\left(\frac{-\omega_{ng} \pm i\Gamma_{ng} \pm \omega}{\gamma_{ng}}\right)
  \right ]$ &
$\frac{i\sqrt{\pi} C_2}{\gamma_{ng}}\left \{ \frac{ 2(\omega_{ng}
\mp i\Gamma_{ng} \mp \omega)}{\gamma_{ng}^2}
W\left(\frac{-\omega_{ng} \pm i\Gamma_{ng} \pm
\omega}{\gamma_{ng}}\right)
  + \frac{2i}{ \sqrt{\pi} \gamma_{ng}}\right \}$
\\
&\\
  \hline
\end{tabular}
\end{center}
 \end{table}

\section{Energy Denominators for Second-Order Processes}

\subsection{Second-Harmonic Generation}

\subsubsection{$\beta \leq 1$}
 \beqa
 \lefteqn{D_{ll}^{IB}(-2\omega; \omega, \omega)  = \frac{i\sqrt{\pi}}{ \gamma_{lg}} \times }
   \nonumber \\
 & &\left\{ \frac{1}{ \omega}
 \left[ W^{(1)}_{\beta_{l}}( - 2\omega)
 - W^{(1)}_{\beta_{l}}( - \omega) 
 + W^{(1)^{*}}_{\beta_{l}}(\omega)
 - W^{(1)^{*}}_{\beta_{l}}(2\omega) \right ] 
+ \frac{ 1}{2(\omega + i\Gamma_{lg})}
 \left [ W^{(1)}_{\beta_{l}}( -\omega)
  - W^{(1)^{*}}_{\beta_{l}}(\omega)
   \right ] \hspace{0.1 in}\right\} \label{Dllshgibbetashort}
 \eeqa

 \beq
 D_{lm}^{IB}(-2\omega; \omega, \omega)  = \frac{-\pi}{\gamma_{lg}\gamma_{mg}}
 \left \{W^{(1)}_{\beta_{l}}( - 2\omega) W^{(1)}_{\beta_{m}}( - \omega)
 +  W^{(1)^{*}}_{\beta_{l}}(\omega)     \left [ W^{(1)^{*}}_{\beta_m}( 2\omega) + W^{(1)}_{\beta_m}( - \omega) \right ]
 \right \}
   \label{Dlmshgibbetashort}
 \eeq

\subsubsection{$\beta = 1$}

\beq
 D_{ll}^{IB}(-2\omega; \omega, \omega)  = \frac{i\sqrt{\pi}}{ \gamma_{lg}} 
 \left\{ \frac{1}{ \omega} \left[ W_{l}( - 2\omega) - W_{l}( - \omega)
 + W^{*}_{l}( \omega) - W^{*}_{l}( 2\omega) \right ]
  + \frac{ 1}{2(\omega + i\Gamma_{lg})} \left [ W_{l}( - \omega) - W^{*}_{l}( \omega) \right ] \hspace{0.1 in}\right\}
   \label{Dllshgibshort}
 \eeq
 \beq
 D_{lm}^{IB}(-2\omega; \omega, \omega)  = \frac{-\pi}{\gamma_{lg}\gamma_{mg}}
\left \{ W_{l}( - 2\omega) W_{m}(- \omega) + W^{*}_{l}( \omega)
\left [  W^{*}_{m}( 2\omega)  + W_{m}(- \omega) \right ] \right \}
   \label{Dlmshgibshort}
 \eeq
\subsection{Linear Electrooptic Effect}
\subsubsection{$\beta \leq 1$}
 \beqa
 D_{ll}^{IB}(-\omega; \omega, 0) & =&
  \frac{i\sqrt{\pi}}{\gamma_{lg}}
  \left\{ \frac{-1}{ \gamma_{lg}}\left[
    W^{(2)^{*}}_{\beta_l}(\omega)
     + W^{(2)}_{\beta_l}( - \omega)
     \right]
 + \frac{2(\omega + i\Gamma_{lg})}{\omega(\omega + 2i\Gamma_{lg})} \left[
 W^{(1)}_{\beta_l}( - \omega) -
 W^{(1)^{*}}_{\beta_l}(\omega)\right]
  \right.
  \nonumber \\
& & \hspace{0.43 in} \left.
 + \frac{  2i\Gamma_{lg}}{(\omega + 2i\Gamma_{lg}) }
 \left[ W^{(1)^{*}}_{\beta_l}(0)
  - W^{(1)}_{\beta_l}(0)
 \right]  \right\}
  \label{Dlleaibbetashort}
 \eeqa
 \beqa
 D_{lm}^{IB}(-\omega; \omega, 0) & =&  \frac{-\pi}{\gamma_{lg}\gamma_{mg}}
 \left\{ W^{(1)}_{\beta_{l}}( - \omega) \left [ W^{(1)}_{\beta_m}(0) + W^{(1)}_{\beta_{m}} (- \omega) \right]
 + W^{(1)^{*}}_{\beta_l}(\omega)\left[ W^{(1)^{*}}_{\beta_m}( \omega) +
 W^{(1)}_{\beta_m}(0)\right]  \right.
 \nonumber \\
 & & \hspace{0.43 in}\left. +  W^{(1)^{*}}_{\beta_l}(0) \left[ W^{(1)^{*}}_{\beta_m}(\omega) - W^{(1)}_{\beta_{m}}( - \omega)\right]
  \right \}
 \label{Dlmeaibbetashort}
 \eeqa
\subsubsection{$\beta = 1$}
 \beqa
  D_{ll}^{IB}(-\omega; \omega, 0) & =&
  \frac{i\sqrt{\pi}}{\gamma_{lg}}
  \left\{ \frac{2}{ \gamma_{lg}^2}\left[
    (\Omega^{*}_{lg} + \omega)W^{*}_{l}( \omega)
      + (\Omega_{lg} - \omega)W_{l}( - \omega)
     \right]  + \frac{4i}{\gamma_{lg}\sqrt{\pi}}
 + \frac{2(\omega + i\Gamma_{lg})}{\omega(\omega + 2i\Gamma_{lg})} \left[
 W_{l}( - \omega) - W^{*}_{l}( \omega)\right] \right.
  \nonumber \\
  & & \hspace{0.43 in} \left.
 + \frac{  2i\Gamma_{lg}}{(\omega + 2i\Gamma_{lg}) }
 \left[ W^{*}_{l}(0)
  - W_{l}(0)
 \right]  \right\}
  \label{Dlleaibshort}
 \eeqa

 \beq
 D_{lm}^{IB}(-\omega; \omega, 0)  =  \frac{-\pi}{\gamma_{lg}\gamma_{mg}}
 \left\{W_{l}( - \omega) \left [ W_{m}(0) + W_{m}(- \omega) \right]  +
  W^{*}_{l}(\omega) \left[ W^{*}_{m}( \omega) + W_{m}(0) \right]
  + W^{*}_{l}(0) \left[ W^{*}_{m}( \omega) - W_{m}(- \omega)\right]
  \right \}
 \label{Dlmeaibshort}
 \eeq
\subsection{Sum and Difference Frequency Generation}
\subsubsection{$\beta \leq 1$}
  \beqa
D_{ll}^{IB}(-(\omega_1 + \omega_2); \omega_1 ,\omega_2)  &=&
  \frac{i\sqrt{\pi}}{ \gamma_{lg}} \left\{
  \frac{1}{\omega_1 + \omega_2 + 2i\Gamma_{lg}}
 \left[ W^{(1)}_{\beta_l}(-\omega_1)  - W^{(1)^{*}}_{\beta_l}(\omega_2)
  +  W^{(1)}_{\beta_l}(-\omega_2) - W^{(1)^{*}}_{\beta_l}(\omega_1)\right] \right.
   \nonumber \\
 & &  +  \frac{1}{\omega_1 }
  \left[ W^{(1)}_{\beta_l}(-\omega_1-\omega_2) - W^{(1)}_{\beta_l}(-\omega_2)
  + W^{(1)^{*}}_{\beta_l}(\omega_1+\omega_2) - W^{(1)^{*}}_{\beta_l}(\omega_2)\right]
  \nonumber \\
  &&\left.+ \frac{1}{\omega_2}
 \left[ W^{(1)}_{\beta_l}(-\omega_1-\omega_2)
   - W^{(1)}_{\beta_l}(-\omega_1)  +
   W^{(1)^{*}}_{\beta_l}(\omega_1+\omega_2)
 - W^{(1)^{*}}_{\beta_l}(\omega_1)\right]
 \right\}
  \label{Dllsdibbetashort}
 \eeqa

 \beqa
D_{lm}^{IB}(-(\omega_1 + \omega_2); \omega_1 ,\omega_2 )  &=&
 \frac{-\pi}{\gamma_{lg}\gamma_{mg}}
\left\{
  W^{(1)}_{\beta_l}(-\omega_1-\omega_2)
  \left[ W^{(1)}_{\beta_m}(-\omega_1) + W^{(1)}_{\beta_m}(-\omega_2) \right]
 \right. \nonumber \\
 & & \hspace{-0.7in} + \left.
 W^{(1)^{*}}_{\beta_l}(\omega_2)
  \left[
  W^{(1)^{*}}_{\beta_m}(\omega_1+\omega_2)
 + W^{(1)}_{\beta_m}(-\omega_1) \right]
   + W^{(1)^{*}}_{\beta_l}(\omega_1)
  \left[
  W^{(1)^{*}}_{\beta_m}(\omega_1+\omega_2)
 + W^{(1)}_{\beta_m}(-\omega_2)
 \right] \right\}
 \label{Dlmsdibbetashort}
 \eeqa
\subsubsection{$\beta = 1$}
  \beqa
 D_{ll}^{IB}(-(\omega_1 + \omega_2); \omega_1 ,\omega_2)  &=&
  \frac{i\sqrt{\pi}}{ \gamma_{lg}}
 \left\{  \frac{1}{\omega_1 }
  \left[  W_{l}( - \omega_1 - \omega_2)   - W_{l}( - \omega_2)
 + W^{*}_{l}( \omega_1 + \omega_2) - W^{*}_{l}( \omega_2 )\right]
 \right.
  \nonumber \\
 & &  + \frac{1}{\omega_2}\left[ W_{l}( - \omega_1 -\omega_2) - W_{l}( - \omega_1) +
   W^{*}_{l}(  \omega_1 + \omega_2) - W^{*}_{l}(  \omega_1)\right]
 \nonumber \\
 & & +  \left.\frac{1}{\omega_1 + \omega_2 + 2i\Gamma_{lg}}
 \left[ W_{l}( - \omega_1 )  - W^{*}_{l}( \omega_2)  +
   W_{l}( - \omega_2)
 - W^{*}_{l}(  \omega_1)\right]
 \right\}
  \label{Dllsdibshort}
 \eeqa

 \beqa
 D_{lm}^{IB}(-(\omega_1 + \omega_2); \omega_1 ,\omega_2 )  &=&
 \frac{-\pi}{\gamma_{lg}\gamma_{mg}}
 \left\{
  W_{l}( - \omega_1 - \omega_2)
  \left[
  W_{m}(- \omega_1 )
 + W_{m}(- \omega_2) \right]\right.
 \nonumber \\
  && \left. +
 W^{*}_{l}( \omega_2)
  \left[
  W^{*}_{m}( \omega_1  + \omega_2)
 + W_{m}(- \omega_1 ) \right]
 + W^{*}_{l}( \omega_1)
  \left[
  W^{*}_{m}( \omega_1 + \omega_2)
 + W_{m}(- \omega_2)
 \right] \right\}
 \label{Dlmsdibshort}
 \eeqa
\twocolumn


\end{document}